\documentclass[sigconf]{acmart}

\usepackage{makecell} 
\usepackage{multirow,enumitem} 
\usepackage{amsfonts} 
\usepackage{mathtools} 
\usepackage{hyperref} 
\usepackage{pgfplots,pgfplotstable}
\usepackage[normalem]{ulem}
\usepackage{subcaption}

\usepackage[linesnumbered,ruled,vlined]{algorithm2e}

\SetCommentSty{mycommfont}

\usepackage{textcomp} 
\usepackage{xcolor} 

\def\BibTeX{{\rm B\kern-.05em{\sc i\kern-.025em b}\kern-.08em
    T\kern-.1667em\lower.7ex\hbox{E}\kern-.125emX}}

\usetikzlibrary{pgfplots.groupplots}
\usetikzlibrary{patterns}
\pgfplotsset{width=0.35\textwidth, compat=1.15}
\setlist[itemize]{noitemsep, topsep=0pt}

\newenvironment{customlegend}[1][]{%
  \begingroup
  \csname pgfplots@init@cleared@structures\endcsname
  \pgfplotsset{#1}%
}{%
  \csname pgfplots@createlegend\endcsname
  \endgroup
}%
\def\addlegendimage{\csname pgfplots@addlegendimage\endcsname}

\AtBeginDocument{%
  \providecommand\BibTeX{{%
    \normalfont B\kern-0.5em{\scshape i\kern-0.25em b}\kern-0.8em\TeX}}}

\copyrightyear{2023}
\acmYear{2023}
\setcopyright{acmlicensed}
\acmConference[CIKM '23]{Proceedings of the 32nd ACM International Conference on Information and Knowledge Management}{October 21--25, 2023}{Birmingham, United Kingdom}
\acmBooktitle{Proceedings of the 32nd ACM International Conference on Information and Knowledge Management (CIKM '23), October 21--25, 2023, Birmingham, United Kingdom}
\acmPrice{15.00}
\acmDOI{10.1145/3583780.3615044}
\acmISBN{979-8-4007-0124-5/23/10}

\newcommand{\m}{\textsf{SAGE}}

\usepackage{pdfrender}

\newcommand*{\boldcheckmark}{%
  \textpdfrender{
    TextRenderingMode=FillStroke,
    LineWidth=.5pt, 
  }{\checkmark}%
}
\begin{document}

\title{{\m}: A Storage-Based Approach for Scalable and Efficient Sparse Generalized Matrix-Matrix Multiplication}


\author{Myung-Hwan Jang}
\authornote{Two first authors have contributed equally to this work.}
\affiliation{%
  \institution{Department of Computer Science\\Hanyang University}
  \city{Seoul}
  \country{Republic of Korea}
}
\email{sugichiin@hanyang.ac.kr}
\orcid{0000-0003-4419-5148}

\author{Yunyong Ko\footnotemark[1]}
\affiliation{%
  \institution{Department of Computer Science\\Hanyang University}
  \city{Seoul}
  \country{Republic of Korea}
}
\email{koyunyong@hanyang.ac.kr}
\orcid{0000-0003-1283-4697}

\author{Hyuck-Moo Gwon}
\affiliation{%
  \institution{Department of Computer Science\\Hanyang University}
  \city{Seoul}
  \country{Republic of Korea}
}
\email{howling6@hanyang.ac.kr}
\orcid{0000-0002-8765-1863}

\author{Ikhyeon Jo}
\affiliation{%
  \institution{Department of Computer Science\\Hanyang University}
  \city{Seoul}
  \country{Republic of Korea}
}
\email{childyouth@hanyang.ac.kr}
\orcid{0000-0003-2245-6220}

\author{Yongjun Park\footnotemark[2]}
\affiliation{%
  \institution{Department of Computer Science\\Yonsei University}
  \city{Seoul}
  \country{Republic of Korea}
}
\email{yongjunpark@yonsei.ac.kr}
\orcid{0000-0003-3725-0380}

\author{Sang-Wook Kim}
\authornote{Corresponding authors.}
\affiliation{%
  \institution{Department of Computer Science\\Hanyang University}
  \city{Seoul}
  \country{Republic of Korea}
}
\email{wook@hanyang.ac.kr}
\orcid{0000-0002-6345-9084}

\renewcommand{\shortauthors}{Jang and Ko, et al.}

\begin{abstract}
Sparse generalized matrix-matrix multiplication (SpGEMM) is a fundamental operation for real-world network analysis.
With the increasing size of real-world networks, 
the single-machine-based SpGEMM approach cannot perform SpGEMM on large-scale networks, exceeding the size of main memory (i.e., \textit{not scalable}).
Although the distributed-system-based approach could handle large-scale SpGEMM based on multiple machines,
it suffers from severe inter-machine communication overhead to aggregate results of multiple machines (i.e., \textit{not efficient}).
To address this dilemma, in this paper, 
we propose a novel \textit{storage-based} SpGEMM approach (\textbf{{\m}}) that stores given networks in storage (e.g., SSD) and loads only the necessary parts of the networks into main memory when they are required for processing via a 3-layer architecture. 
Furthermore, 
we point out three challenges that could degrade the overall performance of {\m} and propose three effective strategies to address them: 
(1) \textit{block-based workload allocation} for balancing workloads across threads, 
(2) \textit{in-memory partial aggregation} for reducing the amount of unnecessarily generated storage-memory I/Os,
and (3) \textit{distribution-aware memory allocation} for preventing unexpected buffer overflows in main memory.
Via extensive evaluation,  
we verify the superiority of {\m} over existing SpGEMM methods in terms of scalability and efficiency.
\end{abstract}



\maketitle

\vspace{-3mm}
\section{Introduction}\label{sec:intro}

Graphs are widely used to model real-world networks, 
where a node represents an object and an edge does the \textit{pair-wise} relationship between two objects.
To analyze real-world networks and discover useful knowledge,
many graph algorithms have been studied~\cite{Sed11,Liu16,Pag99,Kle99,Suz03,Kar98,Bec08,He10,Kep11,Kep16,Kan20,Ham17,Ko21},
where a graph is generally represented as a \textit{dense} matrix (i.e., 2-D array), and each element $e_{i,j}=1$ if an edge exists between nodes $i$ and $j$, and $e_{i,j}=0$ otherwise.
\textit{Generalized matrix-matrix multiplication} (GEMM), one of the key operations in these algorithms, plays a fundamental role to diffuse the node information via network connectivity. 
Recently, with the growing size of real-world networks, 
it has become more crucial to efficiently perform GEMM on large-scale graphs~\cite{Gu17,Zhe16,Aza16}.

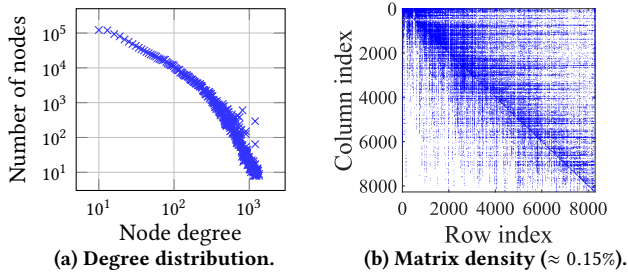
\begin{figure}[t]
	\centering
	\begin{subfigure}[b]{0.245\textwidth}
	\begin{tikzpicture}
	\begin{loglogaxis}[
		height=4cm,
		grid=both,
        major grid style={line width=.2pt,draw=gray!50},
        width=\linewidth,
		xlabel={Node degree},
		xlabel style={font=\normalsize,yshift=3pt,},
		yticklabel style={align=center, yshift = -0pt, font=\footnotesize},
		xticklabel style={font=\footnotesize,},
		ylabel=Number of nodes,
		xtick={1, 10, 100, 1000, 10000},
        ytick={1, 10, 100, 1000, 10000, 100000},
        xmin=5, xmax=3000,
        ymin=3, ymax=500000,
		ymode=log,
        xmode=log,
        log basis x={10},
        log basis y={10},
	]
	\addplot [
    	only marks, 
    	mark=x, 
    	blue!80, 
    	each nth point={3}, 
    	restrict x to domain=0.9:1e4,
    	restrict y to domain=0.9:1e6,
	] table {data/power_wiki.dat};
	\end{loglogaxis}
	\end{tikzpicture}
	\vspace{-3mm}
	\caption{Degree distribution.}
    \end{subfigure}%
    \begin{subfigure}[b]{0.25\textwidth}
        \includegraphics[width=0.88\linewidth]{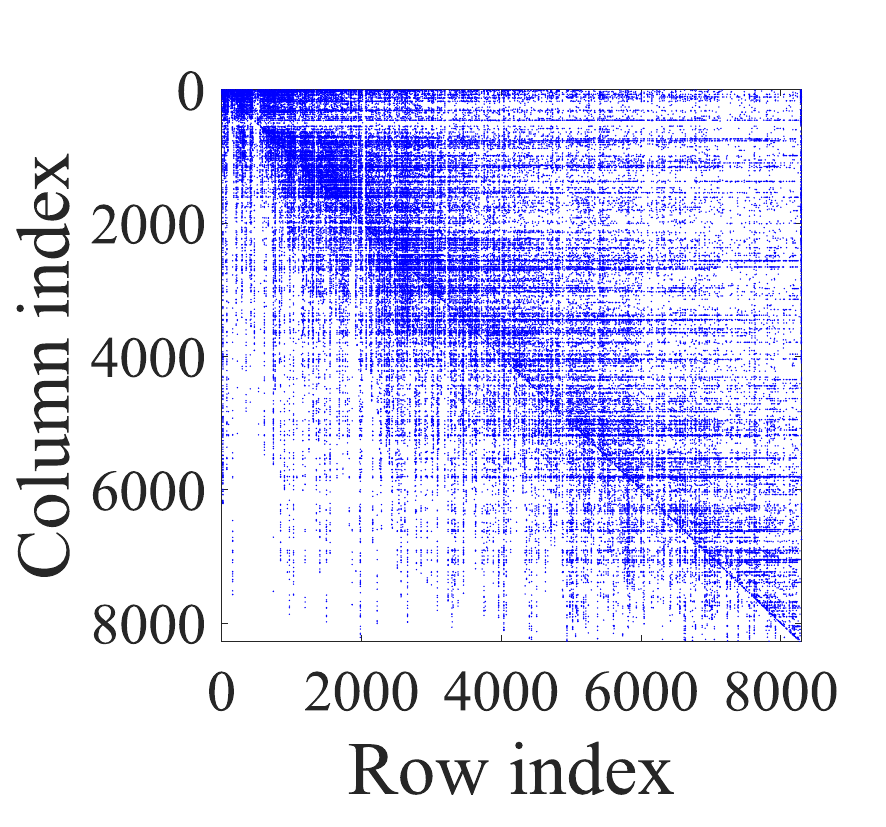}
        \vspace{-3mm}
        \caption{Matrix density ($\approx 0.15\%$).}
    \end{subfigure}
    \vspace{-7mm}
    \caption{(a) Power-law degree distribution and (b) visualized matrix density of real-world graphs.}
    \vspace{-7.5mm}
\label{fig:power}
\end{figure}

Performing GEMM on real-world graphs represented as dense matrices, however, often requires an unnecessarily large amount of space and computational cost~\cite{Nag18,Pal18}.
This is because they tend to follow a \textit{power-law degree distribution}~\cite{Les07} -- i.e., a majority of nodes have only a few edges, while a small number of nodes have a large number of edges -- 
so that the number of non-zero elements (existing edges) is much smaller than that of zero elements (non-existing edges) in the matrix for a real-world graph.
Figure~\ref{fig:power}(a) shows the power-law degree distribution of Wikipedia~\cite{Jo19} and Figure~\ref{fig:power}(b) does the matrix density of Wiki-Vote~\cite{Les10a,Les10b}, where each point represents a non-zero element, respectively.

To address this issue, \textit{sparse} GEMM (\textit{SpGEMM}) methods have been widely studied~\cite{Akb17,Dem20,Aza16,Nag17,Dev17,Gre18,And16,Pal18,Zha20,Dav19}, 
where a graph is represented as a \textit{sparse matrix} containing only existing edges.
On one hand, \textit{single-machine-based} methods~\cite{Akb17,Wan14,Xie19,Nag18} aim to perform SpGEMM efficiently on a single machine by considering the characteristics of real-world graphs, 
where they assume that the sparse matrix can fit in main memory. 
Thus, they are unable to process SpGEMM on large-scale graphs exceeding the size of the main memory (i.e., \textbf{not scalable}).
On the other hand, \textit{distributed-system-based} methods~\cite{Bul12,Aza16,Sel19,Dem20} are able to perform SpGEMM on large-scale graphs, 
exceeding the size of the main memory, by using multiple machines concurrently. 
These methods, however, require a substantial amount of \textit{inter-machine communication} overhead to aggregate the results from multiple machines, which is often more than $50\%$ of the entire process~\cite{Aza16,ko2021depth},
and a lot of \textit{costs and efforts} to maintain complex distributed systems (i.e., \textbf{not efficient}).

To tackle these two limitations together, in this paper,
we propose a novel \textit{storage-based} approach for large-scale SpGEMM,
named \textbf{\uline{S}}torage-b\textbf{\uline{A}}sed approach for Sp\textbf{\uline{GE}}MM (\textbf{{\m}}).
{\m} stores the entire graphs in external storage of a single machine and loads only some parts of the graphs when necessary into the main memory to process them.
For efficient data transfers between storage and main memory,
we design {\m} with a 3-layer architecture, i.e., (1) storage, (2) in-memory, and (3) operation layers, 
where each layer closely interacts with other layers.
Thus, {\m} (1) is able to process large-scale graphs, exceeding the size of the main memory by using \textit{sufficient capacity of external storage} (i.e., \textbf{scalable})
and (2) requires only \textit{intra-machine communication} overhead (storage-memory I/Os), much smaller than the inter-machine communication overhead (i.e., \textbf{efficient}).

To improve the performance of {\m}, 
it is crucial to handle \textit{storage-memory I/Os} efficiently, 
which are inevitably generated due to the limited size of main memory on a single machine. 
To this end, first, we point out three important challenges that could lead to serious performance degradation in {\m}: 
\textbf{(C1) Workload balancing}, i.e., \textit{How to distribute workloads of SpGEMM evenly across multiple threads?},
\textbf{(C2) Intermediate handling}, i.e., \textit{How to handle a large amount of the storage-memory I/Os generated by the intermediate results of SpGEMM?},
and \textbf{(C3) Memory management}, i.e., \textit{How to allocate the main memory space to three buffers to reduce storage-memory I/Os?} (for two input and one output matrices).
Then, we propose three effective strategies to address these challenges: 
(1) \textit{block-based workload allocation} to evenly distribute workloads of SpGEMM into multiple threads for C1,
(2) \textit{in-memory partial aggregation} to reduce the amount of unnecessarily generated storage-memory I/Os for C2,
and (3) \textit{distribution-aware memory allocation} to adjust the proportions of the three buffers, based on the characteristics of real-world graphs for C3.

We note that although there exists several existing storage-based methods~\cite{Kyr12,Roy13,Han13,Zhe15,Jo19},
they focus only on the multiplication between sparse matrix (SpM) and dense vector (DV), i.e., \textit{SpMV} but not \textit{SpGEMM} that we focus on.
More specifically, the existing storage-based SpMV methods handle only ``one" large sparse matrix and small dense vectors ($SpM \times DV = DV$), 
where the dense vectors are much smaller than a sparse matrix ($SpM >> DV$).
On the other hand, {\m} handles ``three" large sparse matrices ($SpM \times SpM = SpM$),
which implies that it is more challenging to efficiently process storage-memory I/Os in the storage-based SpGEMM than the storage-based SpMV.
To the best of our knowledge, this is the first work to successfully perform large-scale SpGEMM by using external storage on a single machine. 
We believe that {\m} could be a practical alternative to researchers and practitioners who aim to analyze large-scale real-world graphs.

The main contributions of this work are summarized as follows:

\begin{itemize}[leftmargin=10pt]
    \item \textbf{New Approach}: We propose a new \textit{storage-based} approach for large-scale SpGEMM, {\m} to address the limitations of existing single-machine-based (\textit{scalability}) and distributed-system-based (\textit{efficiency}) approaches simultaneously.

    \item \textbf{Effective Strategies}: We point out three performance-critical challenges and propose effective strategies to tackle them: (1) \textit{block-based workload allocation}, (2) \textit{in-memory partial aggregation}, and (3) \textit{distribution-aware memory allocation}.
    
    \item \textbf{Extensive Evaluation}: We conduct extensive evaluation, which demonstrates that (1) {\m} is able to perform SpGEMM on large-scale graphs, surpassing the limits of the existing single-machine-based methods, (2) {\m} achieves high SpGEMM performance comparable to or better than the distributed-system-based methods, 
    and (3) each of the proposed strategies of {\m} is effective in improving the performance of {\m}.
\end{itemize}



\vspace{-1mm}
\section{Related Work}\label{sec:related_work}

In this section, we review existing SpGEMM methods.
Table~\ref{table:existing_algorithms} compares {\m} with existing methods in terms of the three performance-critical challenges: (C1) workload balancing (WB), (C2) intermediate handling (IH), and (C3) memory management (MM).

\begin{table}[t!]
\caption{Comparison of existing SpGEMM methods with {\m} based on the three performance-critical challenges}
\vspace{-3mm}
\setlength\tabcolsep{7pt}
\begin{tabular}{cccc}
\toprule
Method & (C1) WB & (C2) IH & (C3) MM \\
\midrule
Intel MKL~\cite{Wan14} & \boldcheckmark &  & \\
KNL-SpGEMM~\cite{Nag18}  & \boldcheckmark &  &  \\
IA-SpGEMM~\cite{Xie19}  & \boldcheckmark &  &  \\

\midrule
SpSUMMA~\cite{Bul12} & \boldcheckmark &  &  \\
Graphulo~\cite{Hut15} & \boldcheckmark &  \boldcheckmark &  \\
Split-3D~\cite{Aza16} & \boldcheckmark &  &  \\
gRRp~\cite{Dem20} & \boldcheckmark & \boldcheckmark & \\
\midrule
\textbf{{\m}} (proposed) & \boldcheckmark & \boldcheckmark & \boldcheckmark \\

\bottomrule
\end{tabular}
\vspace{-5mm}
\label{table:existing_algorithms}
\end{table}

\vspace{1mm}
\noindent
\textbf{Single-machine-based approach.}
A single-machine-based approach~\cite{Wan14,Xie19,Akb17,Nag18} aims to improve the performance of SpGEMM, based on the properties of real-world networks on a single machine (e.g., node degree distribution).
Intel MKL~\cite{Wan14} is an open source library supporting a series of math functions such as SpGEMM and SpMV, which are optimized for Intel CPU architecture.
\citet{Nag18} proposed memory management to keep output results and thread scheduling strategies to resolve the bottlenecks of the entire SpGEMM process.
\citet{Xie19} proposed an input-aware auto-tuning framework for SpGEMM (IA-SpGEMM) to reduce the overhead of memory access and sparse accumulation in SpGEMM.
These single-machine-based methods, however, assume that the entire input matrices can be loaded in the main memory (i.e., in-memory based SpGEMM) because the input matrices have only existing edges in SpGEMM (i.e., sparse matrix).
Therefore, they are not able to process SpGEMM on large-scale graphs exceeding the size of main memory in a single machine. 

There are some works using external storage to process large-scale graphs on a single machine~\cite{Kyr12,Han13,Zhe15,Zhe16,Jo19}.
However, they focus on improving \textit{sparse matrix and dense vector multiplication} (SpMV) rather than SpGEMM that our work focuses on.
Note that we can obtain the exactly same results as SpGEMM by iteratively performing SpMVs in concept; however, 
this way is not suitable for processing SpGEMM since the operations for dense vectors can lead to serious performance degradation due to a large amount of memory usage and computation overhead~\cite{Pal18,Bul12,Aza16}.

\vspace{1mm}
\noindent
\textbf{Distributed-system-based approach.}
A distributed-system-based approach~\cite{Bul12,Hut15,Aza16,Sel19,Dem20} aims to process SpGEMM on large-scale graphs that cannot be loaded on the main memory of a single machine, via multiple machines in a distributed cluster.
This approach splits and stores input matrices into multiple machines, and then processes them in parallel.
\citet{Bul12} proposed SpSUMMA that parallelizes SpGEMM by partitioning input matrices into 2-D grids and distributing them to multiple processors.
\citet{Hut15} proposed a distributed SpGEMM method, Graphulo that considers the \textit{locality} of data stored in a distributed database to reduce the inter-machine communication overhead.
\citet{Aza16} proposed Split-3D that splits input/output matrices into 3-D grids for parallelization of computations and communications.
\citet{Dem20} proposed gRRp, a bipartite graph-based partitioning method for balancing workloads among multiple machines.

Though the distributed-system-based approach is able to perform SpGEMM on large-scale graphs, 
a substantial amount of \textit{inter-machine communication overhead} is required to aggregate the results from multiple machines, 
which is non-trivial (more than $50\%$)~\cite{Aza16,ko2021depth}.
It also requires a lot of \textit{costs and efforts} to maintain the high-performance infrastructure of distributed systems such as fault tolerance and graph partitioning for distributed machines.

\vspace{-1mm}
\section{{\m}: Proposed framework}\label{sec:proposed}

\begin{table}[t]
\small
\caption{Notations and their descriptions}
\vspace{-3mm}
\setlength\tabcolsep{8pt}
\def\arraystretch{0.86} 
\begin{tabular}{ll}
\toprule
 \textbf{Notation} & \textbf{Description}\\
\midrule
$M^{1}_{in}, M^{2}_{in}$ & input matrices of SpGEMM \\
$M_{out}$ & output matrix ($M^{1}_{in}\times M^{2}_{in}$) \\
$M(i,;)$ & $i$-th row of matrix $M$\\
\midrule
$B^{1}_{in}, B^{2}_{in}$ & buffers for input matrices in the main memory\\
$B_{out}$ & buffer for output matrix in the main memory\\
$T^{1}_{buf}, T^{2}_{buf}$ & buffer index tables for input matrices $M^{1}_{in}, M^{2}_{in}$\\
$T^{1}_{obj}, T^{2}_{obj}$ & object index tables for input matrices $M^{1}_{in}, M^{2}_{in}$\\
\bottomrule
\end{tabular}
\label{table:notations}
\vspace{-5mm}
\end{table}

We present a novel storage-based SpGEMM method, named \textbf{\uline{S}}torage-b\textbf{\uline{A}}sed approach for Sp\textbf{\uline{GE}}MM (\textbf{{\m}}). 
First, we describe the notations and the problem statement (Section~\ref{sec:proposed-notation}). 
Then, we present the architecture and core algorithms of {\m} (Section~\ref{sec:proposed-architecture}), and three novel strategies to address performance-critical challenges (Section~\ref{sec:proposed-optimization}).
Finally, we analyze the complexity of {\m} (Section~\ref{sec:proposed-analysis}).

\vspace{-1mm}
\subsection{Notations and Problem Statement}\label{sec:proposed-notation}

\subsubsection{\textbf{Notations}}
Table~\ref{table:notations} shows the notations used in this paper.
We denote two input sparse matrices for SpGEMM as $M^{1}_{in}$ and $M^{2}_{in}$, 
and the output matrix as $M_{out}$ (i.e., $M^{1}_{in}\times M^{2}_{in}$).
In a matrix $M$, $M(i,;)$ is an $i$-th row, and $M(i,j)$ is a $j$-th element in the $i$-th row, 
where the length of each row ($|M(i,;)|$) can vary across rows because $M$ is a sparse matrix.
For clarity, we denote buffers and index tables in the main memory as $B$ and $T$, respectively.
$B^{1}_{in}$ and $B^{2}_{in}$ are the input buffers to hold necessary parts of the input matrices, and $B_{out}$ is the output buffer for the intermediate results.

\vspace{-1mm}
\subsubsection{\textbf{Problem statement}}
\begin{figure}[t]
    \centering
    \includegraphics[width=0.45\textwidth]{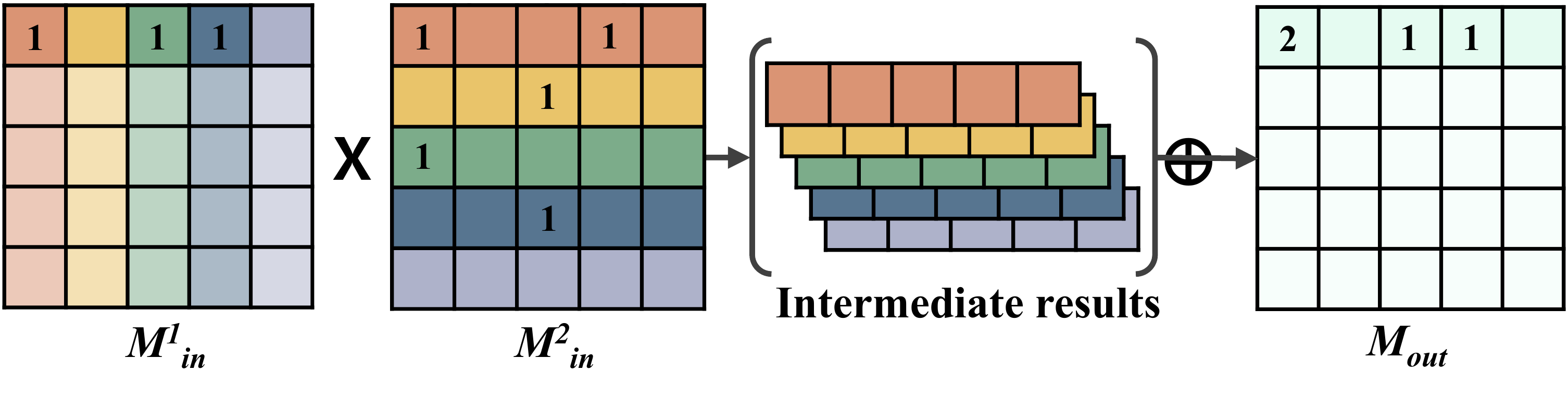}
    \vspace{-4mm}
    \caption{The process of row-wise product.}
    \vspace{-6mm}
\label{fig:spgemm}
\end{figure}

Given two input sparse matrices of $M^{1}_{in}$ and $M^{2}_{in}$, 
{\m} aims to produce the output sparse matrix $M_{out}$ by performing SpGEMM between the two sparse matrices.
When performing SpGEMM, three different types of products can be considered: 
(1) \textit{inner product} (row-by-column product), 
(2) \textit{outer product} (column-by-row product), 
and (3) \textit{row-wise product} (row-by-row product).
Although these three products generate  exactly the same result, $M_{out}$, 
they require different computation overhead and memory space.
Specifically, the row-wise product requires not only much smaller computation overhead (e.g., index-matching) than the inner product, 
but also less memory usage than the outer products~\cite{Sri20,Baek21}.
Thus, we adopt the \textit{row-wise product} for performing SpGEMM in {\m}, following existing algorithms~\cite{Akb17,Wan14,Xie19,Nag18}.
In the row-wise product, the $i$-th row of the output matrix is computed by Eq.~\ref{eq:row-wise},
\begin{equation}
    M_{out}(i,;) = \sum_{j=1}^{n} M^{1}_{in}(i,j) \times M^{2}_{in}(j,;),\label{eq:row-wise}
\end{equation}
where $n$ indicates the number of rows in the input matrices.
Figure~\ref{fig:spgemm} illustrates the process of the row-wise product computing the $i$-th row of the output matrix $M_{out}$ in performing SpGEMM.
First, every $j$-th element of the $i$-th row in the former input matrix, $M^{1}_{in}(i,j)$,
is multiplied by its corresponding $j$-th row in the latter input matrix, $M^{2}_{in}(j,;)$, (highlighted in the same color in Figure~\ref{fig:spgemm}).
Then, the intermediate results are aggregated to compute the final result of the $i$-th row of the output matrix, $M_{out}(i,;)$.
This process is repeated until all rows of the output matrix are computed.

\begin{figure*}[t]
    \centering
    \includegraphics[width=0.85\textwidth]{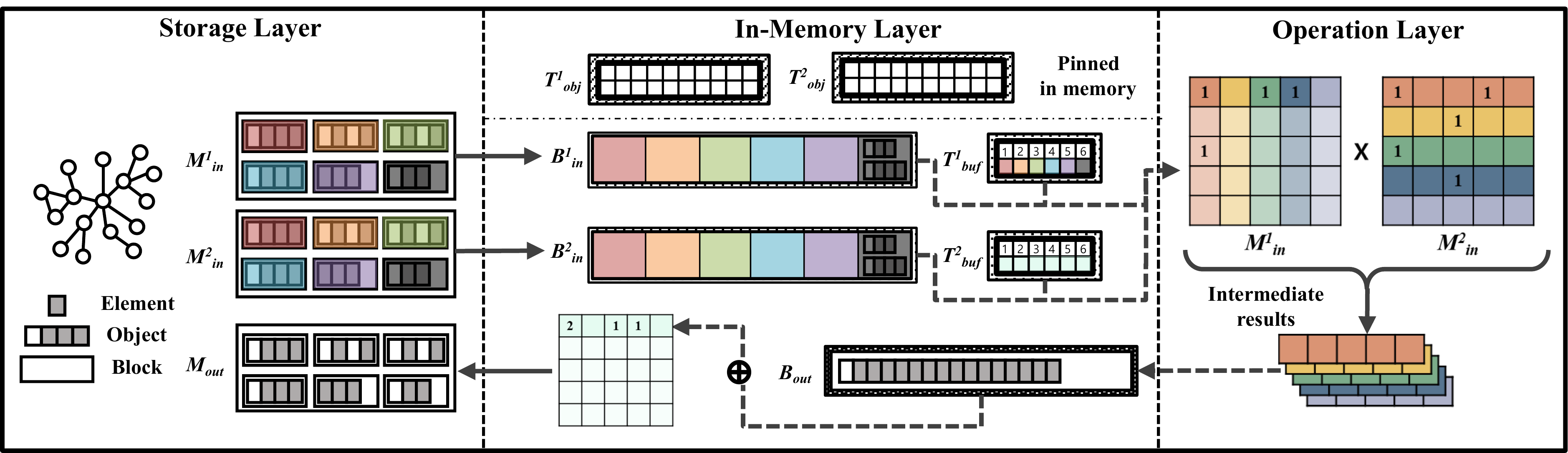}
    \vspace{-3mm}
    \caption{Overview of {\m} with the 3-layer architecture (storage, in-memory, and operation layers).}
    \vspace{-3mm}
    \label{fig:archi}
\end{figure*}

\vspace{-1mm}
\subsection{Architecture and Algorithm}\label{sec:proposed-architecture}
{\m} stores the entire graphs in storage (e.g., SSD), 
loads only the necessary parts of the graphs into the main memory, 
and processes them in parallel by using multiple threads.
Thus, \textit{storage-memory I/Os} are inevitably generated during the process of storage-based SpGEMM,
which could become a \textit{bottleneck} of the overall SpGEMM process.
To handle the inevitable overhead efficiently, in this section,
we propose a novel 3-layer architecture for efficient data transfers between storage and main memory (Section~\ref{sec:proposed-architecture-1}),
and describe how {\m} processes SpGEMM based on the architecture (Section~\ref{sec:proposed-architecture-2}).

\vspace{-1mm}
\subsubsection{\textbf{Architecture}}\label{sec:proposed-architecture-1}
As illustrated in Figure~\ref{fig:archi},
{\m} consists of three layers: (1) storage, (2) in-memory, and (3) operation layers, 
where each layer closely interacts with other layers to efficiently process storage-memory I/Os.

\vspace{1mm}
\noindent
\textbf{Storage layer.}
This layer is in charge of managing the data stored in external storage: two input matrices ($M^{1}_{in}$ and $M^{2}_{in}$), an output matrix ($M_{out}$),
and intermediate results ($M_{out}(i,;)$).
This storage layer handles read/write requests from the in-memory layer.
We define a fixed-size unit standard I/O as a \textit{block} and store the input matrices in a set of blocks.
Each block usually contains multiple objects, each of which stores each row of the input matrices (i.e., a node and its related edges).
In a real-world graph following the power-law degree distribution, 
there are a few nodes with a very high degree, which are too large to be stored in a block.
For the exceptional nodes, we divide and store them in multiple blocks.

\vspace{1mm}
\noindent
\textbf{In-memory layer.}
This layer is in charge of (1) loading blocks required for the next operations into main memory,
(2) aggregating the intermediate results from the operation layer to generate the final result,
and (3) writing the final result to the storage layer.
For this process, we define three buffers and four index tables:
\begin{itemize}[leftmargin=10pt]
    \item Input buffers ($B^{1}_{in}$ and $B^{2}_{in}$): storing rows of the input matrices (i.e., blocks) required for the row-wise product.
    \item Output buffer ($B_{out}$): storing the output result of a row-wise product computed from the operation layer.
    \item Buffer index tables ($T^{1}_{buf}$ and $T^{2}_{buf}$): indicating blocks loaded in the two input buffers of $B^{1}_{in}$ and $B^{2}_{in}$.
    \item Object index tables ($T^{1}_{obj}$ and $T^{2}_{obj}$): indicating the mappings of blocks and objects in external storage.
\end{itemize}

\vspace{1mm}
\noindent
The buffer index tables ($T^{*}_{buf}$) are referred to check whether the required rows are already loaded on the main memory.
If the rows are not on the main memory, 
this layer requests `the block having the required rows' to the storage layer.
In the object index tables ($T^{*}_{obj}$), 
the $i$-th column represents the indices of objects stored in the $i$-th block.
For memory efficiency, 
we sort objects by their indices and store the only two indices of the first and last objects.
The object index tables ($T^{*}_{obj}$) are pinned in the main memory to quickly load the necessary blocks.
Note that the pinned object index tables do not affect the entire performance of {\m}
since it requires only a small amount of memory less than $0.01\%$ of the original graph.

\vspace{1mm}
\noindent
\textbf{Operation layer.}
This layer is in charge of performing actual multiplications using the data in $B^{1}_{in}$ and $B^{2}_{in}$.
This layer is performed as follows:
(1) reading two rows to be multiplied in $B^{1}_{in}$ and $B^{2}_{in}$ by referring to $T^{1}_{buf}$ and $T^{2}_{buf}$,
(2) performing a row-wise product between them,
and (3) writing the result (the intermediate result of $M_{out}(i,;)$) to the output buffer.
This process is repeated until the input matrices are processed.
The intermediate results are aggregated in the in-memory layer to generate the final result $M_{\textit{out}}(i,;)$, and then stored in the external storage.

\begin{algorithm}[b]
\small
\DontPrintSemicolon 
\SetVlineSkip{0.0pt}
\SetInd{0.1em}{1.5em}
\SetKwInOut{Input}{\hspace{0.7em}Input}
\SetKwInOut{Output}{Output}
\Input{Input matrices $M^{1}_{in}$, $M^{2}_{in}$, Object index tables $T^{1}_{obj}$, $T^{2}_{obj}$}
\Output{Output matrix $\mathcal{H}^*$}
\SetKwFunction{FMain}{{\m}}
\SetKwFunction{FSub}{{\sf LoadData}}
\SetKwProg{Pn}{Function}{:}{\KwRet}
\Pn{\FMain{$M^{1}_{in}$, $M^{2}_{in}$, $T^{1}_{obj}$, $T^{2}_{obj}$}}{
    $T^{1}_{buf}, T^{2}_{buf}, M_{\textit{out}}, \text{\sf t\_row} \leftarrow \emptyset$ \\
    \For(\tcp*[f]{processed in multiple threads}){$M^{1}_{in}(i,;) \in M^{1}_{in}$}{
        $\text{\sf LoadData}(i, T^{1}_{obj}, T^{1}_{buf})$ \\
        \For(\tcp*[f]{$i$-th row computation}){$M^{2}_{in}(j,;) \in M^{2}_{in}$} {
            $\text{\sf LoadData}(j, T^{2}_{obj}, T^{2}_{buf})$ \\
            $B_{out} \leftarrow B_{out} \cup \{\text{\sf t\_row}\}, \text{\sf t\_row} \leftarrow  M^{\textit{1}}_{\textit{in}}(i,j) \times M^{2}_{in}(j,;)$ \\
        }
        $M_{out}(i,;) \leftarrow \sum_{\text{\sf tmp} \in B_{out}} \text{\sf tmp}$ \tcp*[f]{result aggregation}
    }
    \Return{$M_{\textit{out}}$}    
}
\Pn{\FSub{$i, T_{obj}, T_{buf}$}}{
    \For{$j=0, 1, ... , |T_{obj}|-1$}{
        $(n_1, n_2) \leftarrow T_{obj}[j]$ \tcp*[f]{first/last nodes in a block} \\
        \If{$n_1 \leq i \leq n_2$}{
            $\text{\sf b\_index} \leftarrow j$ \\
            \textbf{break}
        }
    }
    \If(\tcp*[f]{storage-memory I/O}){$\text{\sf b\_index} \not\in T_{buf}$}{
        $T_{buf} \leftarrow T_{buf} \cup \{\text{\sf b\_index}\}$ \\
    }
    \Return{}    
}
\caption{{\m}: {\sc Storage-based SpGEMM}}\label{algo:sage}
\end{algorithm}

\vspace{-1mm}
\subsubsection{\textbf{Algorithm and performance consideration}}\label{sec:proposed-architecture-2}
Next, let us describe how {\m} performs SpGEMM based on the 3-layer architecture.
Algorithm~\ref{algo:sage} shows the entire process of {\m} (logical view).
Given the two input matrices, which are split and stored in multiple blocks, and their object index tables,
(\textbf{Data loading}) {\m} first loads the two rows to be multiplied into the input buffers ($\text{\sf LoadData}(\cdot)$ in lines 10-18).
Specifically, 
(1) the operation layer requests the two rows to the in-memory layer,
(2) the in-memory layer accesses the input buffer tables to check whether the blocks having the required rows are already in the input buffers,
and (3) the in-memory layer sends the blocks having the required rows to the operation layer if they are already in the input buffers, or otherwise, requests the blocks to the storage layer.
(\textbf{Multiplication}) Then, for the two rows loaded in the main memory, 
{\m} computes the $i$-th row of the output matrix, $M_{out}(i,;)$, by Eq.~\ref{eq:row-wise} (lines 3-8),
where the intermediate results ($\text{\sf t\_row}$) are stored in $B_{out}$ and aggregated to generate the final result $M_{out}(i,;)$ in the in-memory layer.
Finally, the in-memory layer sends (i.e., writes) the final results of the output matrix to the storage layer.

{\m} employs a \textit{multi-thread based row-wise product} to accelerate SpGEMM.
Basically, {\m} assigns each row of the \textit{former} input matrix to each thread, 
rather than the latter input matrix.
Thus, the outer \textbf{for} loop in Algorithm~\ref{algo:sage} (lines 3-8) is performed by multiple threads in parallel.
In this way, {\m} can benefit from the parallelism of SpGEMM since the computation of each thread is \textit{independent} to each other.
Furthermore, to maximize the parallelism, 
we adopt a \textit{block-based workload allocation} strategy to distribute workloads into multiple threads evenly (elaborated in Section~\ref{sec:proposed-optimization-work}).


\begin{figure}[t]
    \centering
    \begin{subfigure}[b]{0.21\textwidth}
        \centering
        \includegraphics[width=0.73\textwidth]{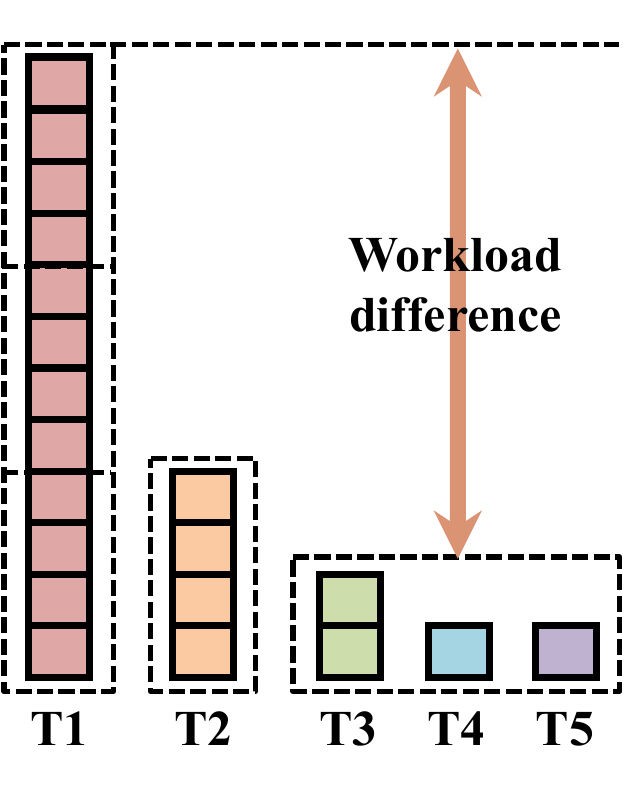}
        \vspace{-3mm}
        \caption{Row-based allocation.}
    \end{subfigure}%
    \begin{subfigure}[b]{0.215\textwidth}
        \centering
        \includegraphics[width=0.77\textwidth]{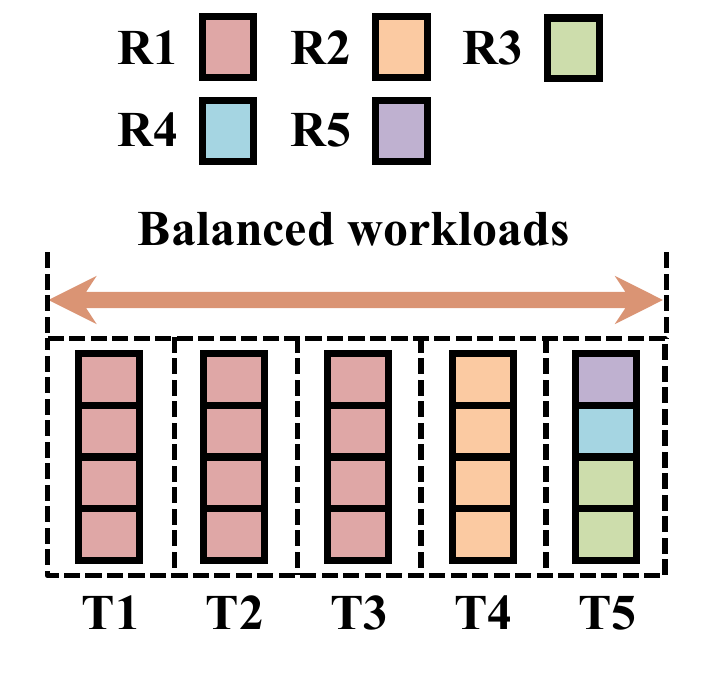}
        \vspace{-3mm}
        \caption{Block-based allocation.}
    \end{subfigure}
    \vspace{-3mm}
    \caption{Comparison of (a) the row-based workload allocation and (b) the block-based workload allocation.}
    \vspace{-4mm}
    \label{fig:opt1-block-based}
\end{figure}

\vspace{-1mm}
\subsection{Challenges and Strategies}\label{sec:proposed-optimization}
We then point out three critical challenges leading to serious performance degradation in {\m} and propose effective strategies to address them: (1) block-based workload allocation, (2) in-memory partial aggregation, and (3) distribution-aware memory allocation.

\vspace{-1mm}
\noindent
\subsubsection{\textbf{Workload balancing}}\label{sec:proposed-optimization-work}
To efficiently perform SpGEMM on large-scale graphs,
existing works generally adopt multi-thread based computation~\cite{Akb17,Wan14,Xie19,Nag18,Bul12,Aza16},
where each thread is in charge of computing different rows (nodes) of the output matrix in parallel (i.e., row-based workload allocation).
Thus, the workload of each thread is decided, depending on the number of elements of each row.
In real-world graphs, however, nodes (rows) have a quite different number of edges (elements) as real-world graphs tend to follow the power-law degree distribution.
It implies that the workloads of threads could be significantly different from each other.
Such workload differences among threads can adversely affect the overall performance of SpGEMM by decreasing the benefit of parallel processing.
Figure~\ref{fig:opt1-block-based}(a) shows an example of row-base workload allocation for five rows with different lengths (R1-R5) stored in five blocks (dotted boxes),
where a large amount of workloads for (R1), stored in three blocks, are assigned on a single thread (T1) (i.e., over-loaded),
while only a small amount of workloads for (R3, R4, and R5), stored in a single block, are split and assigned on three threads (T3-T5) (i.e., under-loaded).
Thus, it is important \textit{to distribute workloads evenly into threads} to maximize the benefit of parallelizing SpGEMM (\textbf{Challenge 1}).


To tackle the challenge of workload balancing, 
we propose a simple yet effective strategy that distributes workloads into multiple threads in a block-based manner (\textbf{block-based workload allocation}).
Specifically, {\m} equipped with the block-based workload allocation assigns ``the same number of blocks'' to each thread, rather than the same number of rows with different numbers of edges, 
thereby distributing the workloads into threads evenly.
Note that this strategy is applied to the outer \textbf{for} loop (line 3) in Algorithm~\ref{algo:sage}, 
where the rows of the former input matrix stored in each block are assigned together to the same thread.
Figure~\ref{fig:opt1-block-based}(b) shows an example of our block-based workload allocation,
where the large row (R1), stored in three blocks, is split and assigned on three threads (T1-T3),
and the small rows (R3-R5), grouped in a single block, are assigned together on a single thread (T5).
As a result, all five threads have the same amount of workloads (one block per thread).

Theoretically, our block-based strategy improves the workload difference among threads, compared to the row-based one, $O(|r|) \rightarrow O(b)$,
where $|r|$ is the \# of elements in a row $r$ and $b$ (fixed) is \# of elements in a block (in general, $|r|_{max} \gg b$).

\begin{figure}[t]
     \centering
     \includegraphics[width=0.385\textwidth]{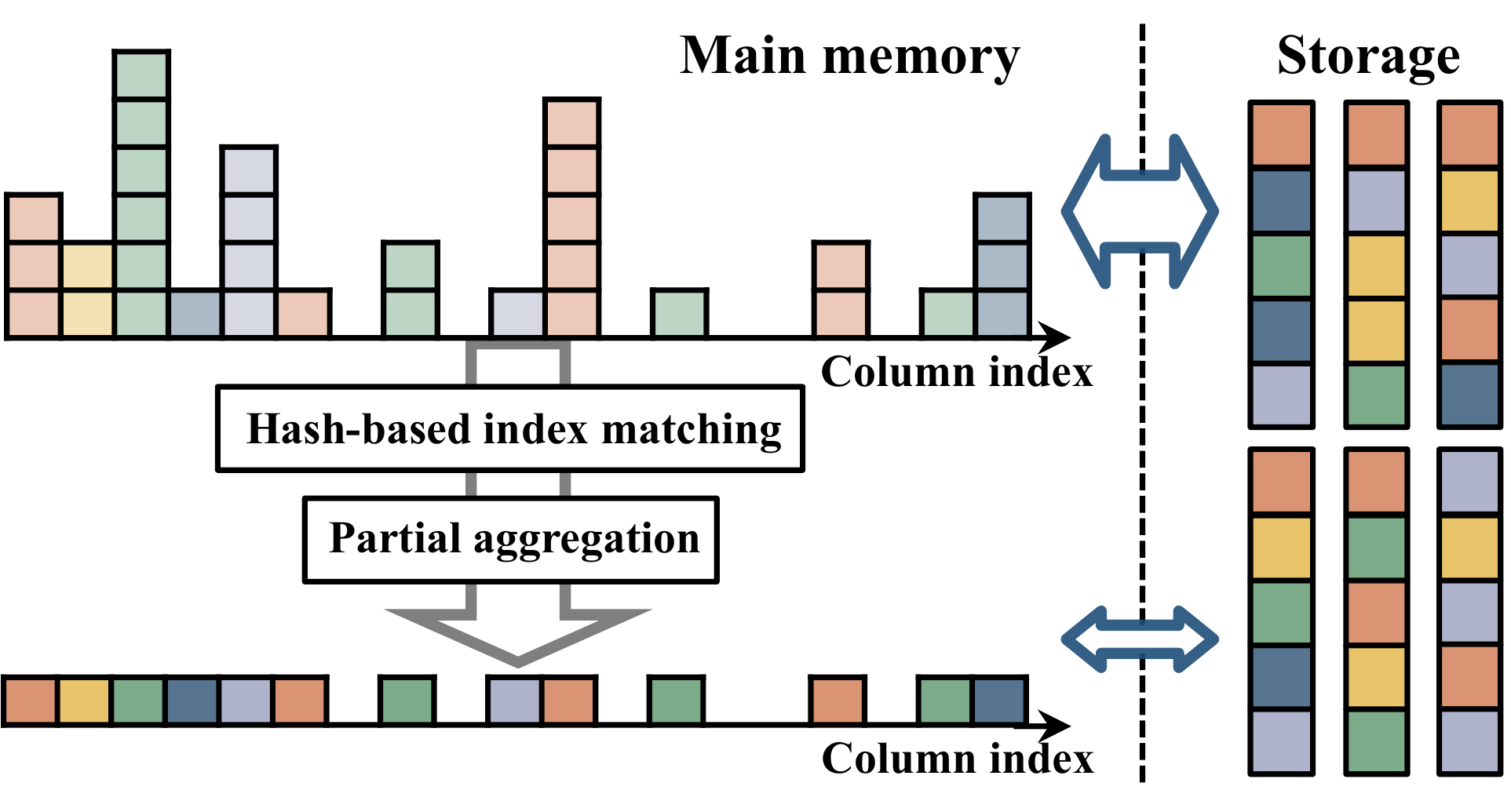}
     \vspace{-3mm}
        \caption{The amount of storage-memory I/Os could be significantly reduced by the in-memory partial aggregation.}
	\vspace{-5mm}
    \label{fig:opt2-inmemory}
\end{figure}

\vspace{-1mm}
\subsubsection{\textbf{Intermediate handling}}\label{sec:proposed-optimization-intermediate}
As described in Section~\ref{sec:proposed-architecture-2},
the $i$-th row of the output matrix ($M_{out}(i,;)$) is computed by aggregating the intermediate results of multiple row-wise products -- the inner \textbf{for} loop (lines 5-8) in Algorithm~\ref{algo:sage}.
For the aggregation, the intermediate results should be loaded on the output buffer.
The output buffer, however, may not be sufficient to store all intermediates due to the limited size of the main memory, 
and it could be overflowed by a large amount of intermediates.
In that case, the intermediates should be flushed into the storage multiple times to process the remaining row-wise products.
They are then loaded back into the main memory to be aggregated in the end once the remaining row-wise products are completed.
In other words, a large amount of storage-memory I/Os could be generated by the intermediates,
which could degrade the performance of SpGEMM significantly.
Therefore, it is critical \textit{to handle a large amount of storage-memory I/Os generated by the intermediates results} (\textbf{Challenge 2}).

To address this challenge, we propose to column-wisely aggregate the intermediate results of each row directly in the output buffer (\textbf{in-memory partial aggregation}).
Figure~\ref{fig:opt2-inmemory} illustrates the effect of our in-memory partial aggregation strategy, 
where the intermediates with the same column indices are aggregated in the main memory before they are flushed to storage.
Accordingly, {\m} is able to reduce a large amount of storage-memory I/Os unnecessarily generated by the intermediates.
Furthermore, we implement \textit{hash-based index matching} in {\m} to quickly find and aggregate the intermediates with the same column index by following~\cite{Nag18}.
However, we have observed that SpGEMM on extremely large-scale graphs may still cause even the partially aggregated intermediates to be overflowed.
In this case, {\m} stores them in storage and aggregate them after finishing all of the row-wise products.

\vspace{-1mm}
\subsubsection{\textbf{Memory management}}\label{sec:proposed-optimization-memory}
{\m} manages three buffers in the main memory (i.e., two input and an output buffers).
Due to the limited size of the main memory,
the amount of storage-memory I/Os can vary depending on the proportions of the three buffers.
For example, (\textbf{Case 1}) if the input buffers are set too small,
the data loaded into the input buffers are processed in only a few iterations and replaced frequently, incurring lots of storage-memory I/Os. 
On the other hand, (\textbf{Case 2}) if the output buffer is set too small,
it is easily overflowed by a small number of intermediate results, 
which also causes storage-memory I/Os frequently.
Therefore, it is crucial \textit{to adjust the proportions of the three buffers} to reduce the amount of storage-memory I/Os
(\textbf{Challenge 3}).

Toward this challenge, 
we propose a practical strategy that adjusts the proportions of the three buffers, based on the characteristics of real-world graphs (\textbf{distribution-aware allocation}).
\uline{First}, {\m} sets the size of the former input buffer $B^{1}_{in}$ as ($b \times t$),
where $b$ is the size of a block and $t$ is the number of threads.
Note that the size of ($b \times t$) is sufficient for the former input buffer to load the blocks that all threads can process simultaneously since {\m} allocates workloads to threads in a \textit{block-based manner}.

\uline{Second}, {\m} determines the size of the output buffer $B_{out}$.
Unfortunately, it is infeasible to predict the exact size of the output row in advance as it depends on data characteristics.
Some previous works~\cite{Wan14,Xie19,Akb17,Nag18} have tried to pre-compute the theoretical maximum size of the intermediate results for each row,
and set the output buffer size as the maximum size in order to prevent the output buffer from being overflowed.
However, it is inappropriate for SpGEMM on real-world graphs following the power-law degree distribution 
since the space required for the intermediate results of a majority of rows is often much smaller than the theoretical maximum size (i.e., inefficient memory usage).
Thus, we argue that the output buffer with a size much smaller than the maximum output size is practically sufficient to process the intermediate results in most cases.
Based on this intuition, we can compute the maximum size of the intermediate results, $max\_row$,
and sets the size of the output buffer as ($\alpha \times max\_row \times t$), considering the degree distribution of real-world graphs (i.e., \textit{distribution-aware allocation}), 
where $\alpha$ is the hyperparameter to adjust the output buffer size.

\uline{Lastly}, {\m} allocates the remaining portion of the main memory to the latter input buffer $B^{2}_{in}$.
Formally, given the capacity of the main memory $C$, a block size $b$, the number of threads $t$, and the hyperparameter $\alpha$,
the main memory is allocated as:
\begin{align}
    {\m}\_mem(C,b,t,\alpha)=
    \begin{cases}
        b \times t, & B^{1}_{in} \\
        C - (|B^{1}_{in}| + |B_{out}|), & B^{2}_{in} \\
        \alpha \times max\_row \times t, & B_{out}.
    \end{cases}
    \label{eq:sage-output}
\end{align}



\subsection{Complexity Analysis}\label{sec:proposed-analysis}
The computational cost of {\m} derives mainly from (1) loading input matrices, (2) row-wise products, (3) aggregating the intermediate results, and (4) storing the final results in the output matrix. 
For brevity, we consider SpGEMM between the two sparse matrices representing the same real-world graph $G=(N,E)$, where $N$ and $E$ is a set of nodes and edges, respectively. 
The input loading requires the time complexity of $O(|N|^2)$ for two input matrices. 
For $|N|$ many rows,
the computational cost of each row-wise products is $O(|N|\times \Tilde{r})$, where $\Tilde{r}$ is the average number of elements in each row, 
where $\Tilde{r} \ll |N|$ since a real-world graph follows the power-law degree distribution.
Thus, the overall complexity of row-wise products is $O(|N|^2\times \Tilde{r})$.
The aggregation of the intermediate results requires $O(|N|^2\times \Tilde{r})$, i.e., (1) hash-based index matching and (2) element-wise aggregation.
Lastly, storing the final results in the output matrix requires the complexity of $O(|N|^2\times \Tilde{r})$.
As a result, the overall time complexity of {\m} is $O(|N|^2\times \Tilde{r})$,
which is comparable to the time complexity of common SpGEMM methods.
We will empirically evaluate the scalability of {\m} with the increasing size of graphs in Section~\ref{sec:eval-eq3-scale}.
\vspace{-1mm}
\section{Experiments}\label{sec:evaluation}

In this section, we comprehensively evaluate {\m} by answering the following evaluation questions (EQs):

\begin{itemize}	[leftmargin=10pt]
    \item \textbf{EQ1}: Does {\m} improve the performance of SpGEMM, compared to the existing single-machine-based methods?
    \item \textbf{EQ2}: Does {\m} improve the performance of SpGEMM, compared to the existing distributed-system-based methods?
    \item \textbf{EQ3}: How does the SpGEMM performance of {\m} scale up with the increasing size of graphs?
    \item \textbf{EQ4}: How effective are the proposed strategies of {\m} in improving the performance of SpGEMM?
\end{itemize}

\vspace{-1.5mm}
\subsection{Experimental Setup}

\begin{table*}[t!]
\caption{Comparison of the single-machine-based approach with {\m} on 18 real-world datasets with various sizes.}
\vspace{-3mm}
\small
\setlength\tabcolsep{5pt} 
\def\arraystretch{0.95} 
\begin{tabular}{c||ccccccccc}
\toprule
Dataset & \textbf{Poisson3D} & \textbf{Enron} & \textbf{Epinions} & \textbf{Sphere} & \textbf{Filter3D} & \textbf{598a} & \textbf{Torso2} & \textbf{Cop20k} & \textbf{Cage12} \\ 
\midrule
\midrule
Intel MKL & \textbf{0.1131} & \textbf{0.5792} & \textbf{0.6343} & \textbf{0.2521} & \textbf{0.3921} & \textbf{0.2660} & \textbf{0.0940} & \textbf{0.4120} & \textbf{0.3669} \\ 
\textbf{{\m}} & 0.3206 & 0.7497 & 0.7001 & 0.5335 & 0.7436 & 0.5626 & 0.3976 & 0.7433 & 0.6331 \\ 
\midrule
Difference (sec.) & 0.2075$\uparrow$ & 0.1705$\uparrow$ & 0.0658$\uparrow$ & 0.2814$\uparrow$ & 0.3515$\uparrow$ & 0.2966$\uparrow$ & 0.3036$\uparrow$ & 0.3313$\uparrow$ & 0.2662$\uparrow$ \\
\midrule
\midrule

Dataset & \textbf{Slashdot} & \textbf{Gowalla} & \textbf{Pokec} & \textbf{Youtube} & \textbf{Livejournal} & \textbf{Wikipedia} & \textbf{UK-2005} & \textbf{SK-2005} & \textbf{Yahoo} \\
\midrule
Intel MKL & 1.9890 & 7.8299 & 12.8732 & \textcolor{red}{N/A, OOM} & \textcolor{red}{N/A, OOM} & \textcolor{red}{N/A, OOM} & \textcolor{red}{N/A, OOM} & \textcolor{red}{N/A, OOM} & \textcolor{red}{N/A, OOM} \\ 
\textbf{{\m}} & \textbf{1.7558} & \textbf{6.772} & \textbf{10.6391} & \textbf{38.734} & \textbf{94.8283} & \textbf{2,276} & \textbf{616} & \textbf{3,709} & \textbf{12,994} \\ 
\midrule
Difference (sec.) & \textcolor{blue}{0.2332$\downarrow$} & \textcolor{blue}{1.0579$\downarrow$} & \textcolor{blue}{2.2341$\downarrow$} & - & - & - & - & - & - \\


\bottomrule
\end{tabular}
\vspace{-2mm}
\label{table:eq1-single}
\end{table*}

\noindent
\textbf{Datasets.}
We use widely used 26 real-world datasets~\cite{Jur14,Kol19} with various sizes, ranging from 10MB at the minimum to 60GB at the maximum, and synthetic datasets.
For the synthetic datasets,
we use the \textbf{R}ecursive \textbf{MAT}rix generator (R-MAT)~\cite{Cha04} to generate synthetic graphs following a power-law degree distribution.
R-MAT takes several parameters: (1) a benchmark type (Graph500, SSCA, or ER), 
(2) the scale of the graph $n$, (3) the edge factor $e$, and (4) the skewness parameters $a,b,c$, and $d$, determining the skewness of the generated graph.
Based on the input parameters, 
R-MAT generates a matrix $M \in \mathbb{R}^{2^{n} \times 2^{n}}$ with $e \times 2^{n}$ elements.
We use the same parameters as used in~\cite{Dem20,Bul12,Aza16}.

\vspace{1mm}
\noindent
\textbf{Evaluation protocol.}
To comprehensively evaluate {\m}, we consider two different types of SpGEMM: 
(1) SpGEMM between ``the same two matrices'' ($M \times M$) and (2) SpGEMM between ``two different matrices'' ($M_1 \times M_2$ or $M \times M^{T}$).
The first type (1) is used in many graph algorithms such as finding all-pairs shortest paths~\cite{Dal07}, self-similarity joins~\cite{He10}, and summarization of sparse datasets~\cite{Ord16},
and the second type (2) is used in collaborative filtering~\cite{Lin03}, triangle counting~\cite{Bec08}, and similarity joins of two different sparse graphs~\cite{Ord16}.
For the evaluation metric,
we use {\it the execution time of SpGEMM} of each method
because the goal of this work is to improve the performance of graph algorithms by accelerating SpGEMM.

\vspace{1mm}
\noindent
\textbf{Competing methods.}
We compare {\m} with the following SpGEMM methods in our experiments.
\begin{itemize}[leftmargin=10pt]
    \item \textbf{Intel MKL}~\cite{Wan14}: the Intel MKL, a state-of-the-art single-machine-based method, is the Intel CPU optimized open-source library supporting a series of math functions such as SpGEMM and SpMV, which was also used in~\cite{Xie19,Akb17,Nag18,Sri20,Pal18}.
    \item \textbf{SpSUMMA}~\cite{Bul12}: SpSUMMA is a distributed-system-based method that parallelizes SpGEMM by partitioning input matrices into 2-D grids and processing them using multiple machines.
    \item \textbf{Graphulo}~\cite{Hut15}: Graphulo stores input matrices in a distributed system exploiting the data locality to reduce the inter-machine communication overhead.
    \item \textbf{gRRp}~\cite{Dem20}: gRRp adopts a bipartite graph-based partitioning method for balancing workloads among multiple machines in a distributed system.
\end{itemize}

\vspace{1mm}
\noindent
\textbf{Implementation details.}
We run all experiments on a single machine equipped with Intel i7-7700K CPU, 64GB main memory, and 4TB SSD as the storage.
We set the number of threads $t$ as 4 to fully utilize all four physical cores in the CPU.
As explained in Section~\ref{sec:proposed-optimization-memory},
{\m} allocates the main memory into three buffers (i.e., two input and one output matrices) by the \textit{distributed-aware allocation} (Eq.~\ref{eq:sage-output}), 
where we set the hyperparameter $\alpha$ as 12.5\%.
We will empirically evaluate the impact of the hyperparameter $\alpha$ on the performance of {\m} in Section~\ref{sec:eval-eq4-ablation}.

\vspace{-1mm}
\subsection{Experimental Results}
\subsubsection{\textbf{Single-machine-based approach (EQ1)}}\label{sec:eval-eq1-single}
We first compare {\m} with the single-machine-based approach, Intel MKL~\cite{Wan14}, in terms of the SpGEMM performance (i.e., execution time).

\vspace{1mm}
\noindent
\textbf{Setup.}
To comprehensively evaluate {\m}, we use 18 real-world datasets with various sizes, 
12 relatively small datasets that can be loaded on the main memory and 6 large datasets that exceed the limit of the main memory capacity. 
Then, we (1) perform SpGEMM using each method on the datasets for 5 times and (2) measure the average execution time of SpGEMM of each method.
Note that we limit the main memory size to 16GB to rigorously evaluate the ability of {\m} to handle the storage-memory I/Os.

\vspace{1mm}
\noindent
\textbf{Results and analysis.}
Table~\ref{table:eq1-single} shows the results. 
\underline{First},
{\m} provides the comparable performance to or even ``better'' than the Intel MKL, the in-memory based method.
Specifically, {\m} (1) finishes SpGEMM slightly slower (only less than 0.4 sec.) than the Intel MKL in 9 datasets which are small enough to load the entire data on the main memory and (2) even outperforms the Intel MKL in Slashdot, Gowalla, and Pokec datasets.
We highlight that these improvements over the Intel MKL are significant,
given that the Intel MKL is an \textit{in-memory based SpGEMM} method loading and processing the entire data on the main memory (i.e., no storage-memory I/Os) and specially designed for the Intel CPU architecture, 
while {\m} has to handle the inherent additional overhead for data transfers between the main memory and external storage as described in Section~\ref{sec:proposed-architecture}.
\underline{Second},
{\m} successfully performs SpGEMM on very large real-world datasets (i.e., from Youtube to Yahoo datasets in Table~\ref{table:eq1-single}) that the Intel MKL could not handle due to the limited capacity of the main memory.

As a result, these results demonstrate that 
{\m} is able not only (1) to perform SpGEMM on very large graphs by utilizing the sufficient capacity of external storage (i.e., \textit{scalable}),
but also (2) to efficiently handle storage-memory I/Os by addressing the three challenges,
as described in Section~\ref{sec:proposed-optimization}.

\begin{table}[t]
\caption{Statistics of synthetic datasets}
\vspace{-3mm}
\small
\setlength\tabcolsep{8pt} 
\def\arraystretch{0.85} 
\begin{tabular}{c|ccc}
\toprule
Benchmark & Scale $n$ & \# of nodes & \# of edges \\
\midrule
\multirow{4}{*}{\makecell{Graph500 \\ (Graphulo and gRRp)}} & 15 & 32.7K & 524K \\ 
 & 16 & 65.6K & 1.04M \\ 
 & 17 & 131K & 2.09M \\ 
 & 18 & 262K & 4.19M \\ 
\midrule
\multirow{4}{*}{\makecell{SSCA  \\ (SpSUMMA)}} & 21 & 2.09M & 16.7M \\ 
 & 22 & 4.19M & 33.5M \\ 
 & 23 & 8.38M & 67.1M \\ 
 & 24 & 16.7M & 134M \\


\bottomrule
\end{tabular}
\vspace{-4mm}
\label{table:eq2-synthetic}
\end{table}

\begin{table*}[t!]
\caption{Comparison of the distributed-system-based methods with {\m} on 13 real-world datasets with various sizes.}
\vspace{-3mm}
\small
\setlength\tabcolsep{3pt} 
\def\arraystretch{1.0} 
\begin{tabular}{c||ccccccccccccc}
\toprule
Dataset & \textbf{Sphere} & \textbf{Filter3D} & \textbf{598a} & \textbf{Torso2} & \textbf{Cage12} & \textbf{144} & \textbf{Wave} & \textbf{Majorbasis} & \textbf{Scircuit} & \textbf{Mac} & \textbf{Offshore} & \textbf{Mario} & \textbf{Tmt\_sym} \\ 
\midrule
\midrule
Graphulo & 7.37 & 12.91 & 9.57 & 4.81 & 13.28 & 11.97 & 11.85 & 8.07 & 6.36 & 4.63 & 18.86 & 9.07 & 13.55 \\ 
gRRp & \underline{1.50} & \underline{3.71} & \underline{1.65} & \underline{0.86} & \underline{2.63} & \underline{2.35} & \underline{1.87} & \underline{1.52} & \underline{1.02} & \underline{1.31} & \underline{3.89} & \underline{1.70} & \underline{3.34} \\ 
\textbf{{\m}} & \textbf{0.36} & \textbf{0.77} & \textbf{0.34} & \textbf{0.11} & \textbf{0.54} & \textbf{0.48} & \textbf{0.41} & \textbf{0.25} & \textbf{0.18} & \textbf{0.25} & \textbf{0.93} & \textbf{0.30} & \textbf{0.62} \\ 
\midrule
Difference (sec.) & \textcolor{blue}{1.14$\downarrow$} & \textcolor{blue}{2.94$\downarrow$} & \textcolor{blue}{1.31$\downarrow$} & \textcolor{blue}{0.75$\downarrow$} & \textcolor{blue}{2.09$\downarrow$} & \textcolor{blue}{1.87$\downarrow$} & \textcolor{blue}{1.46$\downarrow$} & \textcolor{blue}{1.27$\downarrow$} & \textcolor{blue}{0.84$\downarrow$} & \textcolor{blue}{1.06$\downarrow$} & \textcolor{blue}{2.96$\downarrow$} & \textcolor{blue}{1.40$\downarrow$} & \textcolor{blue}{2.72$\downarrow$} \\


\bottomrule
\end{tabular}
\vspace{-3mm}
\label{table:eq2-distributed-real}
\end{table*}

\vspace{-1mm}
\subsubsection{\textbf{Distributed-system-based approach (EQ2)}}\label{sec:eval-eq2-distributed}
In this experiment, we compare {\m} with three state-of-the-art distributed-system-based SpGEMM methods: Graphulo~\cite{Hut15}, gRRp~\cite{Dem20}, and Sparse SUMMA (SpSUMMA)~\cite{Bul12}.
We use their experimental results reported in~\cite{Dem20,Bul12} by following ~\cite{Kyr12,Roy13,Jo19,Maass17,Jun18} since we use the exactly same datasets as in~\cite{Dem20,Bul12} and it is not feasible to construct the identical distributed systems that the existing works used.

\vspace{1mm}
\noindent
\textbf{Setup.}
We use both real-world and synthetic datasets, which were used in~\cite{Dem20,Bul12}.
Specifically, for comparison with Graphulo and gRRp, 
we use 13 real-world and 4 synthetic datasets, generated by Graph500 benchmark with the scale factor varying from 15 to 18,
where we set the edge factor $e=16$ and the skewness parameters $a=0.57$, $b=0.19$, $c=0.19$, and $d=0.05$ by following~\cite{Dem20}.
For comparison with SpSUMMA, we use 4 synthetic datasets, generated by SSCA benchmark with the scale factor varying $n$ from 21 to 24, where we set the edge factor $e=8$ and the skewness parameters $a=0.6$ and $b=c=d=0.4/3$ by following~\cite{Bul12}.
Table~\ref{table:eq2-synthetic} shows the statistics of the synthetic datasets used in this experiment.

Then, we (1) perform SpGEMM using {\m} on the datasets for 5 times and (2) measure the average execution time of SpGEMM of each method.
We note that Graphulo and gRRp have been conducted on the distributed system that consists of 10 machines equipped with 2 Intel Xeon E5-2690 v4 processors connected via the DGS-3120-24TC Ethernet~\cite{Dem20} and 
SpSUMMA has been conducted on the Franklin XT4 system,
where each machine is equipped with a quad-core AMD Opteron processor and connected with each other via a 6.4 GB/s network~\cite{Bul12}.

\vspace{1mm}
\noindent
\textbf{Results and analysis.}
Table~\ref{table:eq2-distributed-real} shows the results of Graphulo, gRRp, and {\m} on 13 real-world datasets.
The results reveal that {\m} consistently outperforms the state-of-the-art distributed-system-based methods in all real-world datasets.
Specifically, {\m} finishes the SpGEMM by up to $7.7\times$ faster and gRRp, the best performer among competitors.
Whereas, Graphulo provides the poor SpGEMM performance on real-world graphs (e.g., by up to $43\times$ slower than {\m} in Torso2 dataset).
This is because Graphulo, which adopts the outer product, suffers from the significant overhead of indexing and aggregating a large amount of the intermediate results of outer-product based SpGEMM as described in Section~\ref{sec:proposed-notation}.
Although gRRp provides the performance of SpGEMM much better than Graphulo,
{\m} still significantly outperforms gRRp.
We argue that these performance improvement of {\m} over Graphulo and gRRp is significant,
considering they perform SpGEMM on the high-performance distributed system that consists of 10 machines~\cite{Dem20}.
Thus, these results imply that {\m} is able to efficiently handle large-scale SpGEMM in a single machine, requiring only intra-machine communication overhead (i.e., storage-memory I/Os), much smaller than the inter-machine communication overhead.

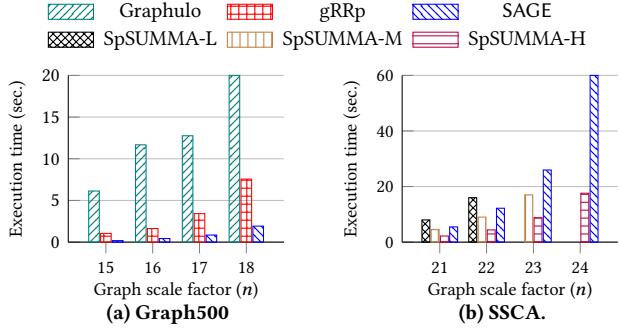
\begin{figure}[t]
\centering
\begin{tikzpicture}
\centering
\small
\begin{customlegend}[legend columns=3,legend style={align=center,draw=none,column sep=0.5ex},
        legend entries={Graphulo, gRRp, {\m}, SpSUMMA-L, SpSUMMA-M, SpSUMMA-H}]
        \addlegendimage{draw=teal, pattern = north east lines, area legend, pattern color = teal}
        \addlegendimage{draw=red, pattern = grid, area legend, pattern color = red}
        \addlegendimage{draw=blue, pattern = north west lines, area legend, pattern color = blue}
        \addlegendimage{draw=black, pattern = crosshatch, area legend, pattern color = black}
        \addlegendimage{draw=brown, pattern = vertical lines, area legend, pattern color = brown}
        \addlegendimage{draw=purple, pattern = horizontal lines, area legend, pattern color = purple}
        \end{customlegend}
\end{tikzpicture}
\begin{subfigure}[b]{0.25\textwidth}
\begin{tikzpicture}
\pgfplotsset{
    xticklabel={\tick},
    scaled x ticks=false,
    plot coordinates/math parser=false,
}
\begin{axis}[
    height=3.8cm,
    ybar=0.5pt,
    bar width=4pt,
    width=1.0\linewidth,
    axis x line*=bottom,
    axis y line*=left,
    log origin=infty,
    yminorticks=false,
    ymajorgrids,
    enlarge x limits={abs=0.5cm},
    xlabel style={font=\footnotesize},
    ylabel style={font=\footnotesize},
    xticklabel style={font=\footnotesize},
    yticklabel style={font=\footnotesize},
    ylabel={Execution time (sec.)},
    ylabel style={align=center},
    xlabel={Graph scale factor ($n$)},
    xlabel style={yshift=3pt,},
    ymin=0, ymax=20,
    ytick={0,5,10,15,20,25},
	symbolic x coords={15,16,17,18},
	after end axis/.code={
        },
	xtick={15,16,17,18},
	nodes near coords align=vertical,
	clip=false,
	legend style={at={(0.5,1.38)},anchor=north, draw=},
	legend cell align=center,
	legend columns=3,
 	legend style={font=\small},
]
\addplot[font=\scriptsize, teal, pattern = north east lines, area legend, pattern color = teal, 	point meta=y,	visualization depends on=rawy\as\rawy,     ]
	coordinates {
(15,6.13)
(16,11.67)
(17,12.76)
(18,20) 
};
\addplot[font=\scriptsize, red, pattern = grid, area legend, pattern color = red, 	point meta=y,	visualization depends on=rawy\as\rawy,     ]
	coordinates {
(15,1.05)
(16,1.63)
(17,3.42)
(18,7.56)
};
\addplot[font=\scriptsize, blue, pattern = north west lines, area legend, pattern color = blue, 	point meta=y,	visualization depends on=rawy\as\rawy,     ]
	coordinates {
(15,0.18)
(16,0.43)
(17,0.85)
(18,1.92)
};
\end{axis}
\end{tikzpicture}
\vspace{-3mm}
\caption{Graph500}
\end{subfigure}%
\begin{subfigure}[b]{0.25\textwidth}
\begin{tikzpicture}
\pgfplotsset{
    xticklabel={\tick},
    scaled x ticks=false,
    plot coordinates/math parser=false,
}
\begin{axis}[
    normalsize,
    height=3.8cm,
    ybar=0.5pt,
    bar width=3pt,
    width=1.0\linewidth,
    axis x line*=bottom,
    axis y line*=left,
    log origin=infty,
    yminorticks=false,
    ymajorgrids,
    enlarge x limits={abs=0.5cm},
    ylabel={Execution time (sec.)},
    ylabel style={align=center},
    xlabel={Graph scale factor ($n$)},
    xlabel style={yshift=3pt,},
    xlabel style={font=\footnotesize},
    ylabel style={font=\footnotesize},
    xticklabel style={font=\footnotesize},
    yticklabel style={font=\footnotesize},
    ymin=0, ymax=60,
    ytick={0,20,40,60},
    yticklabels={0,20,40,60},
	symbolic x coords={21,22,23,24},
	after end axis/.code={
        },
	xtick={21,22,23,24},
	nodes near coords align=vertical,
	clip=false,
	legend style={at={(0.5,1.38)}, anchor=north, draw=},
	legend cell align=center,
	legend columns=2,
 	legend style={font=\small},
]
\addplot[font=\scriptsize, black, pattern = crosshatch, area legend, pattern color = black, 	point meta=y,	visualization depends on=rawy\as\rawy,     ]
	coordinates {
(21,8)
(22,16)
};
\addplot[font=\scriptsize, brown, pattern = vertical lines, area legend, pattern color = brown, 	point meta=y,	visualization depends on=rawy\as\rawy,     ]
	coordinates {
(21,4.5)
(22,9)
(23,17)
};
\addplot[font=\scriptsize, purple, pattern = horizontal lines, area legend, pattern color = purple, 	point meta=y,	visualization depends on=rawy\as\rawy,     ]
	coordinates {
(21,2.2)
(22,4.4)
(23,8.8)
(24,17.6)
};
\addplot[font=\scriptsize, blue, pattern = north west lines, area legend, pattern color = blue, 	point meta=y,	visualization depends on=rawy\as\rawy,     ]
	coordinates {
(21,5.48)
(22,12.20)
(23,25.96)
(24,60) 
};
\end{axis}
\end{tikzpicture}
\vspace{-3mm}
\caption{SSCA.}
\end{subfigure}%
\vspace{-3mm}
\caption{Comparison of the distributed-system-based methods with {\m} on synthetic datasets.}
\vspace{-5.5mm}
\label{fig:eq2-distributed-synthetic}
\end{figure}

Figure~\ref{fig:eq2-distributed-synthetic} shows the results on synthetic datasets, generated by two benchmarks (Graph500 and SSCA),
where the $x$-axis represents the graph scale factor $n$ of each synthetic dataset and the $y$-axis represents the execution time (sec.).
Similar to the results on real-wold datasets,
Figure~\ref{fig:eq2-distributed-synthetic}(a) shows that {\m} significantly outperforms the two distributed-system-based methods, Graphulo and gRRp, across all graph scales.
Moreover, as shown in Figure~\ref{fig:eq2-distributed-synthetic}(b)\footnote{\citet{Bul12} used three different number of cores in its system with 3 (SpSUMMA-L), 4 (SpSUMMA-M), and 9 (SpSUMMA-H) machines, respectively.}, 
{\m} successfully performs SpGEMM on very large-scale graphs (the scale factor $n\geq 23$) that SpSUMMA-L and SpSUMMA-M fail to handle in a single machine.
As a result, These results demonstrate that {\m} is able to process large-scale real-world graphs successfully with only limited computing resources of a single machine, rather than those of costly distributed-systems (i.e., cost-effective).

\input{figures/eq3-scalability}

\vspace{-1mm}
\subsubsection{\textbf{Scalability (EQ3).}}\label{sec:eval-eq3-scale}
In this experiment, we evaluate the scalability of {\m} with the increasing sizes of graphs.

\vspace{1mm}
\noindent
\textbf{Setup.}
We use synthetic datasets, generated by two different benchmarks (Graph500 and ER)~\cite{Aza16} with various graph scales $n$ ranging from 18 to 28, where the scale factor $n$ decides the number of nodes in a generated graph (i.e., $|N|=2^n$).
For each generated graph, we also consider two different edge factors to control the number of edges (i.e., one for a high-degree graph and the other for a low-degree graph).
Specifically, we set the edge factor for the high-degree and low-degree graphs as $1.5\times e$ and $0.5\times e$, respectively\footnote{We could not show the results of high-degree graphs with graph scale 28 since R-MAT generator could not generate them due to the integer overflow.}.
Then, we perform SpGEMM (type (2)) on the generated graphs for 5 times and measure the average execution time (sec.).

\vspace{1mm}
\noindent
\textbf{Results and analysis.}
Figure~\ref{fig:eq3-scalability} shows the results,
where the $x$-axis represents the scale factor $n$, the $y$-axis represents the execution time (sec. in log-scale), and the red dotted vertical line represents the limits of existing single-machine-based methods.
We observe that the execution time of {\m} tends to increase by 4.5 times as the scale factor increases by 2 (i.e., a graph gets $4\times$ larger).
This result demonstrates that the SpGEMM performance of {\m} successfully scales up to very large-scale graphs, exceeding the capacity of the main memory (red dotted vertical line),
and {\m} provides (almost) linear scalability with the increasing sizes of graphs.


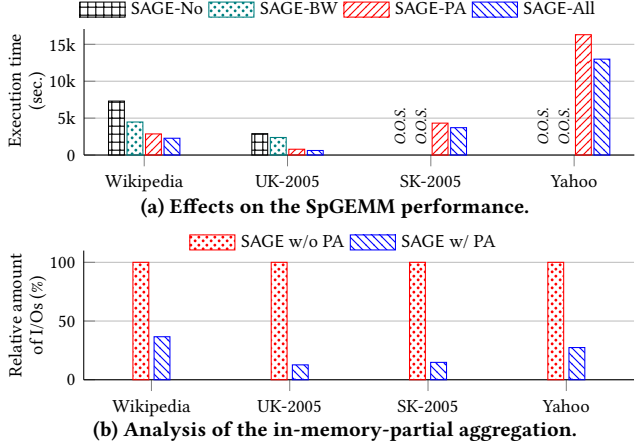
\begin{figure}[t]
\centering
\begin{subfigure}[b]{0.5\textwidth}
\begin{tikzpicture}
\pgfplotsset{
    xticklabel={\tick},
    scaled x ticks=false,
    plot coordinates/math parser=false,
}
\begin{axis}[
    normalsize,
    height=3.3cm,
    width=1.0\linewidth,
    axis x line*=center,
    axis y line*=left,
    ybar=1pt,
    bar width=6pt,
    ymajorgrids,
    enlarge x limits={abs=0.8cm},
    ylabel={Execution time\\(sec.)},
    ylabel style={align=center,yshift=-5pt},
    scaled y ticks = false,
    ymin=0, ymax=17500,
    ytick={0,5000,10000,15000},
    yticklabels={0,5k,10k,15k,},
    xlabel style={yshift=3pt,font=\footnotesize},
            ylabel style={align=center,font=\footnotesize},
            xticklabel style={font=\footnotesize},
            yticklabel style={font=\footnotesize},
	symbolic x coords={Wikipedia, UK-2005, SK-2005, Yahoo},
	xtick={Wikipedia,UK-2005,SK-2005,Yahoo},
	xticklabel style={},
	nodes near coords align=vertical,
	clip=false,
	legend style={at={(0.465,1.275)},anchor=north, draw=none},
	legend cell align=center,
	legend columns=4,
	legend style={font=\footnotesize},
]
\addplot[font=\scriptsize, black, pattern = grid, area legend, pattern color = black, 	point meta=y,	visualization depends on=rawy\as\rawy,]     
	coordinates {
(Wikipedia,7310)
(UK-2005,2896)
};\addlegendentry{{\m}-No}
\addplot[font=\scriptsize, teal, pattern = crosshatch dots, area legend, pattern color = teal, 	point meta=y,	visualization depends on=rawy\as\rawy,]     
	coordinates {
(Wikipedia,4476)
(UK-2005,2379)
};\addlegendentry{{\m}-BW}
\addplot[font=\scriptsize, red, pattern = north east lines, area legend, pattern color = red, 	point meta=y,	visualization depends on=rawy\as\rawy,]     
	coordinates {
(Wikipedia,2862)
(UK-2005,787)
(SK-2005,4324)
(Yahoo,16304)
};\addlegendentry{{\m}-PA}
\addplot[font=\scriptsize, blue, pattern = north west lines, area legend, pattern color = blue, 	point meta=y,	visualization depends on=rawy\as\rawy,]     
	coordinates {
(Wikipedia,2276)
(UK-2005,616)
(SK-2005,3709)
(Yahoo,12994)
};\addlegendentry{{\m}-All}
\end{axis}
\node[font=\scriptsize,color=black,rotate=90] at (4.20,0.38) {\textit{O.O.S.}};
\node[font=\scriptsize,color=black,rotate=90] at (4.48,0.38) {\textit{O.O.S.}};
\node[font=\scriptsize,color=black,rotate=90] at (6.10,0.38) {\textit{O.O.S.}};
\node[font=\scriptsize,color=black,rotate=90] at (6.38,0.38) {\textit{O.O.S.}};
\end{tikzpicture}
\vspace{-2.75mm}
\caption{Effects on the SpGEMM performance.}
\end{subfigure}
\vspace{-1mm}
\begin{subfigure}[b]{0.5\textwidth}
\begin{tikzpicture}
\pgfplotsset{
    xticklabel={\tick},
    scaled x ticks=false,
    plot coordinates/math parser=false,
}
\begin{axis}[
    normalsize,
    height=3.3cm,
    width=1.0\linewidth,
    axis x line*=bottom,
    axis y line*=left,
    ybar,
    bar width=6pt,
    ymajorgrids,
    enlarge x limits={abs=0.9cm},
    ylabel={Relative amount\\of I/Os (\%)},
    ylabel style={align=center,yshift=-5pt},
    ymin=0, ymax=110,
    scaled y ticks = false,
    ytick={0,50,100},
    yticklabels={0,50,100},
    xlabel style={yshift=3pt,font=\footnotesize},
            ylabel style={align=center,font=\footnotesize},
            xticklabel style={font=\footnotesize},
            yticklabel style={font=\footnotesize},
	symbolic x coords={Wikipedia, UK-2005, SK-2005, Yahoo},
	after end axis/.code={
        },
	xtick={Wikipedia, UK-2005, SK-2005, Yahoo},
	nodes near coords align=vertical,
	clip=false,
	legend style={at={(0.475,1.21)},anchor=north, draw=none},
	legend cell align=center,
	legend columns=2,
	legend style={font=\footnotesize},
]
\addplot[font=\small, red, pattern = crosshatch dots, area legend, pattern color = red, 	point meta=y,	visualization depends on=rawy\as\rawy,]
	coordinates {
(Wikipedia, 100)
(UK-2005,100)
(SK-2005,100)
(Yahoo,100)
};\addlegendentry{{\m} w/o PA}
\addplot[font=\small, blue, pattern = north west lines, area legend, pattern color = blue, 	point meta=y,	visualization depends on=rawy\as\rawy,]
	coordinates {
(Wikipedia,36.641)
(UK-2005,12.681)
(SK-2005,14.8)
(Yahoo,27.417)
};\addlegendentry{{\m} w/ PA}
\end{axis}
\end{tikzpicture}
\vspace{-3mm}
\caption{Analysis of the in-memory-partial aggregation.}
\end{subfigure}
\vspace{-6mm}
\caption{Effects of our proposed strategies on {\m}.}
\vspace{-7mm}
\label{fig:eq4-ablation}
\end{figure}
 

\vspace{-1mm}
\subsubsection{\textbf{Ablation study (EQ4).}}\label{sec:eval-eq4-ablation}
Lastly, in this section, we evaluate the effectiveness of the proposed strategies of {\m} individually.

\vspace{1mm}
\noindent
\textbf{Setup (1).}
We first evaluate the first two strategies: (1) block-based workload allocation and (2) in-memory partial aggregation.
To rigorously evaluate the two strategies, 
we (1) use four largest real-world datasets (i.e., Wikipedia, UK-2005, SK-2005, and Yahoo datasets~\cite{Jo19}), which could not be loaded into main memory, 
and (2) limit the size of the main memory to 2GB to incur a large amount of storage-memory I/Os.
We compare the following four versions of {\m}:
\begin{itemize}	[leftmargin=10pt]
    \item {\m}-No: a baseline without any optimizations
    \item {\m}-BW: the version with the block-based workload allocation
    \item {\m}-PA: the version with the in-memory partial aggregation
    \item {\m}-All: the final version with both optimizations
\end{itemize}
We perform SpGEMM (type (1)) using each version and measure the execution time (sec.).
For this experiment, we set the hyperparameter $\alpha=12.5\%$ for the main memory buffer allocation.

\vspace{1mm}
\noindent
\textbf{Results and analysis (1).}
As shown in Figure~\ref{fig:eq4-ablation}(a),
each of the proposed strategy is effective in improving the performance of SpGEMM and {\m}-All consistently provides the best performance in all datasets (up to $4.7\times$ over {\m}-no).
Also, for the two largest datasets (i.e., SK-2005 and Yahoo), 
{\m}-no and {\m}-BW fail to perform SpGEMM because the size of intermediate results even exceeds the capacity of the 4TB external storage, where `O.O.S.' indicates `out of storage'.
Interestingly, {\m}-PA consistently outperforms {\m}-BW in all cases, 
which implies that, in case of large-scale SpGEMM, handling storage-memory I/Os could be more critical than balancing workloads,
in particular when the main memory size of a single machine is limited.

For more analysis of the in-memory partial aggregation,
we measure the amount of storage-memory I/Os generated by the intermediate results during SpGEMM of {\m} with/without the in-memory partial aggregation ({\m} w/ PA and {\m} w/o PA in Figure~\ref{fig:eq4-ablation}(b)).
Figure~\ref{fig:eq4-ablation}(b) shows the relative amount of storage-memory I/Os,
where our in-memory partial aggregation significantly reduces the amount of storage-memory I/Os (up to $87\%$ in UK-2005).
This result demonstrates that in performing storage-based SpGEMM on large-scale graphs,
a substantial amount of storage-memory I/Os could be generated; thus, it is crucial to efficiently handle these I/Os in order to achieve good performance, as we claimed in Section~\ref{sec:proposed-optimization}.


\vspace{1mm}
\noindent
\textbf{Setup (2).}
We also evaluate our distribution-aware memory allocation.
As described in in Section~\ref{sec:proposed-optimization-memory},
we set (1) the former input buffer $B^{1}_{in}= b \times t$, where $b$ is the size of a block (1MB) and $t$ is the number of threads;
and (2) the output buffer $B_{out}= \alpha \times max\_row \times t$, where $max\_row$ here indicates the maximum size of the final results for each row.
Then, we perform SpGEMM with varying (1) the numbers of threads (2, 4, and 8), (2) the size of $B^{1}_{in}$ (1MB to 16MB), and (3) the hyperparameter $\alpha$ (6.25\% to 50\%).

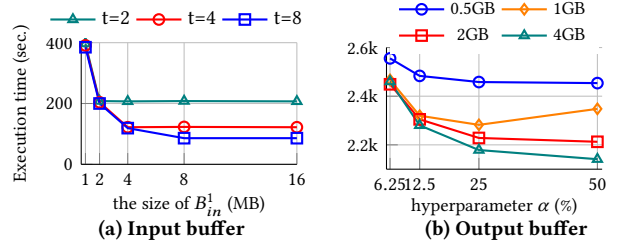
\begin{figure}[t]
\centering
\begin{subfigure}[b]{0.25\textwidth}
\begin{tikzpicture}
\pgfplotsset{
    scaled x ticks=false,
    plot coordinates/math parser=false,
}
\begin{axis}[
    normalsize,
    height=3.2cm,
    width=1.0\textwidth,
    axis x line*=bottom,
    axis y line*=left,
    log origin=infty,
    grid=both,
    xlabel={the size of $B^{1}_{in}$ (MB)},
    title style={font=\normalsize, yshift=-82pt},
    ylabel={Execution time (sec.)},
    ylabel style={align=center},
    ymin=0, ymax=400,
    ytick={0,200,400},
    xmin=0.7, xmax=16,
    xlabel style={yshift=3pt,font=\footnotesize},
            ylabel style={align=center,font=\footnotesize},
            xticklabel style={font=\footnotesize},
            yticklabel style={font=\footnotesize},
	after end axis/.code={
        },
	xtick={1,2,4,8,16,32},
	nodes near coords align=vertical,
	clip=false,
    legend style={at={(0.45,1.35)},anchor=north,draw=none},
	legend cell align=center,
	legend columns=3,
	legend style={font=\small},
]
\addplot[font=\footnotesize, teal, mark=triangle, mark options={mark size=2pt}, line width=0.8pt, 	point meta=y,	visualization depends on=rawy\as\rawy,     ]
	coordinates {
(1,397)
(2,208)
(4,207)
(8,208)
(16,207)
};\addlegendentry{t=2}
\addplot[font=\footnotesize, red, mark=o,mark options={mark size=2pt}, line width=0.8pt, 	point meta=y,	visualization depends on=rawy\as\rawy,     ]
	coordinates {
(1,391)
(2,204)
(4,122)
(8,123)
(16,122)
};\addlegendentry{t=4}
\addplot[font=\footnotesize, blue, mark=square, mark options={mark size=2pt}, line width=0.8pt,	point meta=y,	visualization depends on=rawy\as\rawy,     ]
	coordinates {
(1,385)
(2,200)
(4,119)
(8,86)
(16,86)
};\addlegendentry{t=8}
\end{axis}
\end{tikzpicture}
\vspace{-3mm}
\caption{Input buffer}
\end{subfigure}%
\begin{subfigure}[b]{0.25\textwidth}
\begin{tikzpicture}
\pgfplotsset{
    scaled x ticks=false,
    plot coordinates/math parser=false,
}
\begin{axis}[
    normalsize,
    height=3.2cm,
    width=1.0\textwidth,
    axis x line*=bottom,
    axis y line*=left,
    log origin=infty,
    grid=both,
    title style={font=\normalsize, yshift=-82pt},
    ylabel style={align=center},
    xlabel={hyperparameter $\alpha$ (\%)},
    ymin=2100, ymax=2600,
    scaled y ticks = false,
    ytick={2200,2400,2600},
    yticklabels={2.2k,2.4k,2.6k},
    xmin=5.5, xmax=51,
        xlabel style={yshift=3pt,font=\footnotesize},
            ylabel style={align=center,font=\footnotesize},
            xticklabel style={font=\footnotesize},
            yticklabel style={font=\footnotesize},
	xtick={6.25,12.5,25,50,100},
	nodes near coords align=vertical,
	clip=false,
    legend style={at={(0.5,1.47)},anchor=north,font=\footnotesize,draw=none},
	legend cell align=center,
	legend columns=2,
]
\addplot[font=\footnotesize, blue, mark=o, mark options={mark size=2pt}, line width=0.8pt, 	point meta=y,	visualization depends on=rawy\as\rawy,     ]
	coordinates {
(6.25,2555.400251)
(12.5,2483.464212)
(25,2458.195139)
(50,2453.81492)
};\addlegendentry{0.5GB}
\addplot[font=\footnotesize, orange, mark=diamond, mark options={mark size=2pt}, line width=0.8pt, 	point meta=y,	visualization depends on=rawy\as\rawy,     ]
	coordinates {
(6.25,2466.920254)
(12.5,2319.768401)
(25,2282.189392)
(50,2348.003254)
};\addlegendentry{1GB}
\addplot[font=\footnotesize, red, mark=square, mark options={mark size=2pt}, line width=0.8pt, 	point meta=y,	visualization depends on=rawy\as\rawy,     ]
	coordinates {
(6.25,2448.94336)
(12.5,2304.847382)
(25,2228.34132)
(50,2212.613742)
};\addlegendentry{2GB}
\addplot[font=\footnotesize, teal, mark=triangle, mark options={mark size=2pt}, line width=0.8pt, 	point meta=y,	visualization depends on=rawy\as\rawy,     ]
	coordinates {
(6.25,2460.259902)
(12.5,2280.299921)
(25,2178.648899)
(50,2141.031732)
};\addlegendentry{4GB}
\end{axis}
\end{tikzpicture}
\vspace{-3mm}
\caption{Output buffer}
\end{subfigure}
\vspace{-7mm}
\caption{Effects of distribution-aware memory allocation.}
\vspace{-4.5mm}
\label{fig:eq4-memory}
\end{figure}

\vspace{1mm}
\noindent
\textbf{Results and analysis (2).}
Figure~\ref{fig:eq4-memory}(a) shows that the performance of {\m} gets improved as the number of threads and the size of $B^{1}_{in}$ increase, 
where the performance improvement converges at a specific point where the size of $B^{1}_{in}$ is equal to $b \times t$,
which implies that the buffer size is \textit{sufficient to load the blocks for that all threads can process} simultaneously, 
as explained in Section~\ref{sec:proposed-optimization-memory}.
Figure~\ref{fig:eq4-memory}(b) shows the performance of {\m} with respect to the size of main memory and the hyperparameter $\alpha$.
The performance of {\m} tends to be improved as $\alpha$ and the size of the main memory increase,
but the performance improvement gets smaller as $\alpha$ increases,
which implies that about $12.5 \sim 25\%$ of the maximum size is enough for $B_{out}$ to process the intermediate results of a majority of rows.
As a result, we believe that our distribution-aware memory allocation is effective and practical in large-scale SpGEMM.

\section{Conclusions}\label{sec:conclusion}
In this paper, we point out the limitations of previous approaches for SpGEMM:
(i) the single-machine-based approach cannot handle large-scale graphs surpassing the capacity of the main memory (i.e., \textbf{not scalable}),
and (ii) the distributed-system-based approach requires a substantial amount of inter-machine communication overhead (i.e., \textbf{not efficient}).
To address both challenges of scalability and efficiency,
we propose a novel storage-based approach to SpGEMM (\textbf{{\m}}) with the 3-layer architecture to efficiently handle storage-memory I/Os.
We further identify three important challenges that could cause serious performance degradation in storage-based SpGEMM and propose three effective strategies to address them: 
(1) block-based workload allocation, (2) in-memory partial aggregation, and (3) distribution-aware memory allocation.
Through comprehensive evaluation,
we demonstrate the superiority of {\m} in terms of (1) \textit{scalability}, (2) \textit{efficiency}, and (3) \textit{effectiveness}.


\section{Acknowledgments}
\vspace{0.5mm}
This is a joint work between Samsung Electronics Co., Ltd and
Hanyang University. This work was supported by Institute of Information \& communications Technology Planning \& Evaluation (IITP) grant funded by the Korea government (MSIT) (No.2022-0-00352 and No.RS-2022-00155586).

\clearpage
\balance
\bibliographystyle{ACM-Reference-Format}
\bibliography{ACMconf}


\end{document}